\documentclass[preprint2]{aastex} 
\usepackage{amsmath} 

\newcommand{\bea}{\begin{eqnarray}} 
\newcommand{\eea}{\end{eqnarray}} 

\newcommand{\nn}{\nonumber} 
\newcommand{\vek}[1]{\boldsymbol{#1}}

\renewcommand{\hat}{ { }} 

\slugcomment{}

\shorttitle{Spin of primary black hole in OJ287} 
\shortauthors{Valtonen et al.}

\begin{document}

\title{Measuring the spin of the primary black hole in OJ287}

\author{M.~J.Valtonen\altaffilmark{1,9}, S.~Mikkola\altaffilmark{1}, 
D.~Merritt\altaffilmark{2}, 
A.~Gopakumar \altaffilmark{3,8}, 
H.~J.~Lehto\altaffilmark{1}, T.~Hyv\"onen\altaffilmark{1}, 
H.~Rampadarath\altaffilmark{4,5}, R.~Saunders\altaffilmark{6}, 
M.~Basta\altaffilmark{7} and R.~Hudec\altaffilmark{7,10}} 

\affil{$^1$Tuorla Observatory, Department of Physics and Astronomy, University of Turku, 
    21500 Piikki\"o, Finland} 
\affil{$^2$ Centre for Computational Relativity and Gravitation, Rochester Institute of Technology, 78 Lomb Memorial Drive, Rochester, NY 14623, USA} 
\affil{$^3$Tata Institute of Fundamental Research, Mumbai 400005,
India}
\affil{$^4$Joint Institute for VLBI in Europe (JIVE), Postbus 2, 7990 AA Dwingeloo, The
Netherlands}
\affil{$^5$Leiden Observatory, Leiden University, P.O. Box 9513, NL-2300 RA  Leiden, The Netherlands} 
\affil{$^6$Department of Physics, University of the West Indies, St. Augustine,Trinidad \& Tobago}
\affil{$^7$ Astronomical Institute, Academy of Sciences, Fricova 298, 25165 Ondrejov, Czech Republic} 
\affil{$^8$Theoretisch-Physikalisches Institut,
Friedrich-Schiller-Universit\"at Jena,\\
Max-Wien-Platz 1,
07743 Jena, Germany}
\affil{$^9$Helsinki Institute of Physics, FIN-00014 University of Helsinki, Finland}
\affil{$^{10}$ Czech Technical University in Prague, Faculty of Electrical Engineering,\\
 Technicka 2, 166 27 Praha 6, Czech Republic} 
  
\begin{abstract} 
The compact binary system in OJ287 is modelled to contain a spinning primary black hole with an accretion disk
and a non-spinning secondary black hole. Using Post Newtonian (PN) accurate equations that include 
2.5PN accurate non-spinning contributions,  the leading order general relativistic and classical 
spin-orbit terms, 
the orbit of the binary black hole in  OJ287 is calculated and as expected it depends on the  spin 
of the primary black hole. 
Using the orbital solution, the specific times when the orbit of the secondary 
crosses the accretion disk of the primary are 
evaluated such that the record of observed outbursts from 1913 up to 2007 
is reproduced. 
The timings of the outbursts are quite sensitive to the spin value. 
In order to reproduce all the known outbursts, including a newly discovered one in 1957, 
the Kerr parameter of the primary has to be $0.28 \pm 0.08$. 
 The quadrupole-moment contributions to the equations of motion allow us to constrain 
the `no-hair' parameter to be $1.0\:\pm\:0.3$ where 0.3 is the one sigma error.
This supports the `black hole no-hair theorem' within the achievable precision.

It should be possible to test the present estimate 
in 2015 when the next outburst is due. 
The timing of the 2015 outburst is 
a strong function of the spin: if the spin is 0.36 of the maximal value allowed in general relativity, 
the outburst begins in early November 2015, 
while the same event starts in the end of January 2016 if the spin is 0.2.  
\end{abstract} 


\section{Introduction} 

The BL Lacertae object OJ287 is known to have a quasiperiodic pattern of outbursts at 12 year intervals
\citep{sil88,val08b}. Until now, the system has been modelled as a binary black hole consisting of 
a non-spinning secondary black hole orbiting a more massive non-spinning primary black hole in an eccentric 
orbit having a periodicity 
of about 12 years \citep{leh96}. 
Further, the double peak structure in the light curve of OJ287, with the two peaks
separated by 1 - 2 years, is interpreted as the double impact of 
the secondary black hole on the accretion disk of the primary \citep{leh96}. 
The model has been successful in predicting future outbursts: 
the predictions for the beginning of 1994, 1995 and 2005 outbursts 
were correct within one to two weeks. The prediction for the 1994 outburst was 1994.65 
based on the average outburst interval \citep{sil88}. 
The outburst had already begun when OJ287 was first observed during the fall 1994 observing season at 1994.68; its arrival
on time confirmed the basic periodicity \citep{sil96a,val97}. The probable time of the beginning of this outburst was at
1994.62 based on correlation comparison with light curves of other outbursts. 
Subsequently, an astrophysical model was constructed with impacts on a rigid 
accretion disk and delays of outbursts relative to the disk crossing; the 
model predicted the start of the next outburst at 1995.87  \citep{leh96}. 
The observations gave the best timing at 1995.845 \citep{sil96b,val97}. 
For the 2005 outburst, 
it was necessary to consider the bending of the accretion disk caused by 
the secondary \citep{sun96,sun97}; when combined with the earlier model 
for the radiation burst delay \citep{leh96} the beginning of the 2005 outburst 
was expected at 2005.74. The best determination from observations 
gives 2005.76 \citep{val06a,val08a}. It confirmed the need for relativistic 
precession since without the precession the outburst would have been a year later. 
Finally, the gravitational radiation energy loss was included in the prediction 
for the next outburst at 2007.70 \citep{val07,val08}. 
Recently, the correctness of the prediction within an accuracy of one day was demonstrated \citep{val08b}. 
The model that does not incorporate the gravitational radiation reaction effect is clearly not tenable: 
it predicted the outburst three weeks later. 
The probability of the four major outbursts happening at the predicted times by chance is negligible, 
and the predictive capability of the model requires general relativistic description for the orbit of the binary black hole. 

We do not include the dynamical friction from hitting the accretion disk since the directly affected disk mass is only a few hundred solar mass, too small in comparison with the secondary black hole mass of over $10^{8} M_{\odot}$ to give an observable signal.

In the present study, we allow the primary black hole to spin in the model and probe  its dominant consequences. 
The additional parameters required to describe the dynamics of such a spinning 
binary black hole model for OJ287 can be constrained 
by recently measured historical outbursts as well as the more recently observed outbursts. 
Needless to say, all of these outbursts were not employed while constructing the 
non-spinning binary black hole model for OJ287 \citep{val08b}. 
Recall that the model without spin effects requires six well timed outbursts to provide a unique orbit and 
the usually employed outbursts are those which occurred in the years 
1913, 1947, 1973, 1983, 1984 and 2005. 
The resulting model also explains 1995 and 2007 
outbursts as mentioned above and there is no conflict with the observations 
at any other expected outburst times
 which have not been observed with adequate precision.
Recently, a major outburst that occurred in 1957 
was identified in the historical record \citep{val06b,ram07}. 
 It turns out that this identification allows one to constrain the spinning
binary black hole model for OJ287.
The next expected outburst in 2015 is also spin sensitive and may 
provide an improved measurement of the spin. 

In the literature we have found several measurements of black hole spins.
Genzel et al. (2003) estimated the spin of the Milky Way central black
hole as $ 0.52 \pm 0.2$ times the maximum possible.
\citet{BrennemanReynolds} estimated a spin parameter of
$0.989^{+0.009}_{-0.002}$ for the black hole in the active galaxy
MCG-6-30-15 using relativistically-broadened spectral features from
the inner accretion disk.
There are other estimates 
for stellar mass black holes 
\citep{Liuetal, McClintocketal, Shafeeetal, Gouetal} with resulting estimates varying between $0.65\: {\rm and}\: 1$. 

We begin by reporting the observations from the historical records 
and then explain how we describe the PN-accurate binary orbit. 
Finally, we report 
the results from the timing model with different spin values and present our conclusions.

\section{Observations} 

The Harvard plate stacks contains about 500,000 glass photographic plates and 
as such it is the largest historical plate collection in the world, 
representing about 15\% of all archival photographic plates \citep{Hudec2006}. 
The plates in the Harvard stacks were exposed between years 1885 and 1993 
(with a gap in 1953 - 1968; so called Menzel gap) and 
therefore they provide a possibility to gather optical data as old as 120 years. 
Plates taken with the same telescope belong to one plate series. 
There are several such plate series at Harvard plate stacks. With the use of several such series, 
two of the authors (R.H. \& M.B.) have managed to obtain more than 500 newly measured historical data points for the
blazar OJ287 
and about 3,000 data points for other 8 blazars during their stay at Harvard plate stacks 
in January-February, 2007. In this paper we present the results of the measurements 
on 9 plates of the AC series, measurements covering the outburst of OJ287 in the year 1913. 
After conversion of our estimates to the standard B band, 
we obtain 9 new data points to the historical light curve of the 1913 outburst. 
The estimates of the brightness of the new data points were 
done with the naked eye equipped with a magnifying glass and 
with the help of nearby constant comparison stars of known magnitudes. 
We call this method the modified Argelander method. 
In Table \ref{comparisonstars} we give the comparison stars used for the estimates of the 
data points of the 1913 outburst. Our comparison stars are 
stars 1, 2 and 7 of Craine \citep{Craine77} (hereafter C77). 
\begin{table}[h]
\caption{Comparison stars used for the estimates of the brightness of OJ 287 around the year 1913. The B and V magnitudes
adopted in this paper are the magnitudes given in C77.
\label{comparisonstars}}
\begin{tabular}{lcccl}
ID & RA        & DEC      & B     & V    \\
1 & 08:54:29.5 & 20:10:56 & 12.19 & 11.52\\
2 & 08:55:15.6 & 20:09:44 & 13.49 & 12.80\\
3 & 08:55:19.0 & 20:00:54 & 14.13 & 13.39
\end{tabular}
\end{table}

The Sonneberg Observatory has the second largest archive with over 300,000 photographic plates covering eight
decades beginning in 1924 \citep{kro93,kro05}. The plates were taken by the Sonneberg Sky and Field Patrols. The Field
Patrol monitored 80 fields along or near the northern Milky Way using astrographs and a Schmidt camera at limiting
magnitudes up to 18. The Sky Patrol records the entire northern sky in two colours (red and blue sensitive plates) with 14
short-focus cameras at limiting magnitude of 14 - 15 \citep{kro93,kro05}. Scanning of the plates began in 1990. Thanks to
Sonneberg Observatory, we obtained digitised photographic plates from the plate archives for the period 
1957 - 1994. The plates were both blue and red sensitive, but here we describe 
only results from 223 blue sensitive plates, as there exist more reliable comparison star sequences in blue.
The plates were scanned using the 
HP Scanjet 7400c \citep{kro05}. The general data reduction method is described 
by \citet{inn04} and the Windows based program Maxim DL was used to obtain magnitudes. 
The magnitudes were transformed to standard B system as described by \citep{kro93}. 
Thirteen nonvariable stars within 5 degrees of OJ287 were measured for comparison purposes. 
They were used to obtain for each plate the transformation equation from instrumental 
magnitude to B magnitude. Using the comparison stars, the rms error was found to vary 
from 0.12 to 0.20 magnitudes from plate to plate. Some of this is suspected to arise 
from intrinsic variability of three of the comparison stars. Altogether we estimate 
the rms error of OJ287 measurements as 0.15 magnitudes in B. The B magnitudes were 
transformed to V by using B - V = 0.45 \citep{tak94}. Further, the V-magnitudes were 
converted to the energy flux by $F = 13.18 \times 10^{(-0.4(V - 13.6))}$ where F is in mJy \citep{bes79}. 

In the interval 1957 - 1978 we found 5 outburst events exceeding 29 mJy at 1957, 1959, 1971, 1972 and 1973. 
All except the 1957 outburst are previously known and described elsewhere \citep{leh96,val06b}.   
We have altogether 16 measurements for this season, out of which 3 are new (for older measurements, see 
Valtonen et al. 2006b)

\begin{figure}[ht] 
\epsscale{1.0} 
\plotone{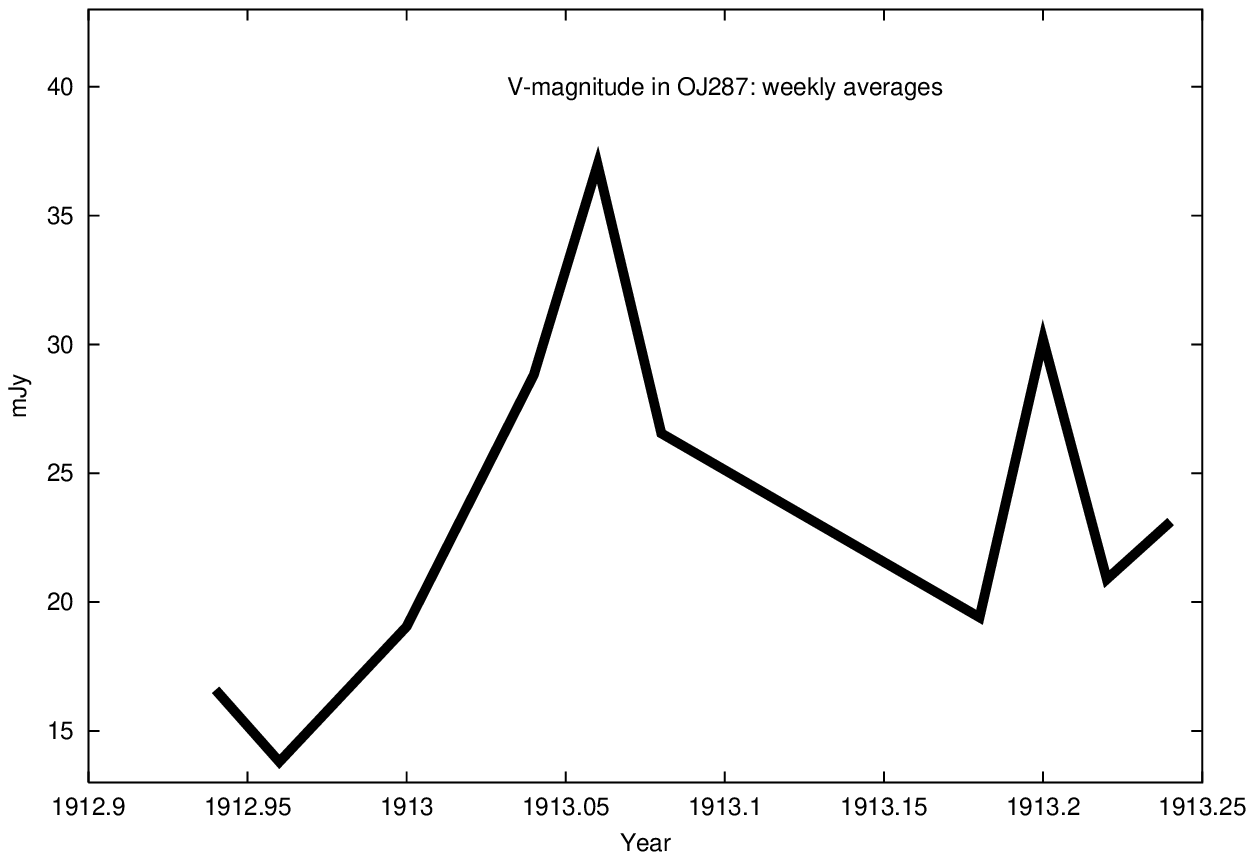} 
\caption{The observation of the brightness of OJ287 in the 1912/3 season: weekly averages. 
The starting time of the outburst is estimated to be 
$1912.98\pm0.02$\label{fig0}} 
\end{figure}

\begin{figure}[ht] 
\epsscale{1.0} 
\plotone{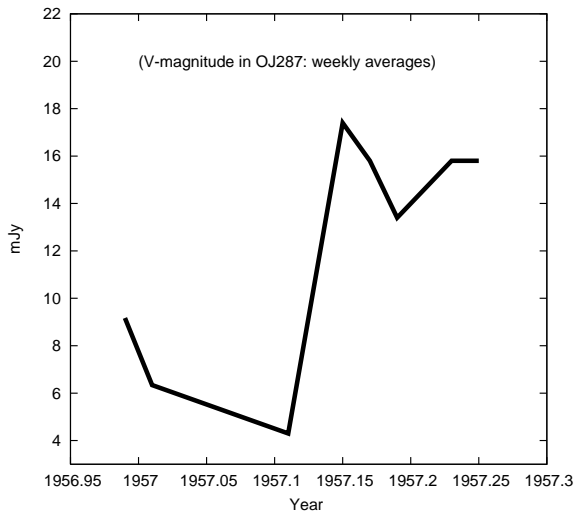} 
\caption{The observation of the brightness of OJ287 in the 1956/7 season: weekly averages. 
The starting time of the outburst is estimated to be $1957.08\pm0.03$
\label{fig1}}
\end{figure}

In Table \ref{newmeasurements2}
we present the newly measured data of the 
light curve of OJ287. 
The estimates of error are based on the quality of the plate. 
The data are also plotted in Figs \ref{fig0} and \ref{fig1} as weekly
averages.
\begin{table}[ht] 
\caption{New measurements ($*=$ bad plate)\label{newmeasurements2}} 
\begin{tabular}{ccccr} 
  yr   &    B  &   V   & error est.&     mJy\\ 

1912.94&  14.00&  13.55&    0.3    &    13.80\\ 

1912.94&  13.60&  13.15&    0.5    &    19.95\\ 

1912.96&  14.00&  13.55&    0.3    &    13.80\\ 

1913.00&  13.65&  13.20&    0.5    &    19.05\\ 

1913.04 & 13.20& 12.75 & 0.3  &    28.84\\ 

1913.08 & 13.15& 12.70 & 0.3  &   30.19\\ 

1913.09 & 13.45& 13.00 & 0.3  &   22.90\\ 

1913.20 & 13.15& 12.70 & 0.4 * &  30.19\\ 

1913.9 & 14.15&  13.70 & 0.4 * &  12.02\\ 



1957.149&  13.31&  12.86&   0.15      &         26.06\\ 

1957.172&  13.22&  12.88&   0.15      &         25.58 \\ 

1957.180&  13.18&  12.73&   0.15      &         29.37\\

\end{tabular} 
\end{table} 

\section{PN-accurate orbital description present in the model} 

  We invoke the 2.5PN-accurate orbital dynamics that includes the leading order 
general relativistic, classical spin-orbit and radiation reaction effects 
for describing the temporal evolution of a binary black hole \citep{kid95}. 
In the present model, where only the primary black hole is spinning, 
the 2.5PN-accurate inspiral dynamics of the binary black hole is fully 
provided by the following differential equations that describe the relative acceleration of the binary 
and the precessional motion for the spin of the primary black hole. 
The 2.5PN-accurate equations of motion can be written schematically as 
\begin{eqnarray} 
\ddot { {\vek x}} \equiv 
\frac{d^2  {\vek x}} { dt^2} &=& 
\ddot { {\vek x}}_{0} + \ddot { {\vek x}}_{1PN} \nonumber 
+ \ddot { {\vek x}}_{SO}+\ddot { {\vek x }}_{Q}\\ 
&& + \ddot { {\vek x}}_{2PN} +  \ddot { {\vek x}}_{2.5PN} \,,  
\end{eqnarray} 
where ${\vek x} = {\vek x}_1 - {\vek x}_2 $ stands for the 
center-of-mass 
relative separation vector between the black holes with masses $m_1$ and $m_2$ and 
$  \ddot { {\vek x}}_{0}  $ represents the Newtonian acceleration given by  
$ \ddot { {\vek x}}_{0} = -\frac{ G\, m}{ r^3 } \, {\vek x} $; $m= m_1 + m_2$ and $ r = | {\vek x} |$. 
The PN contributions occurring at the conservative 1PN, 2PN and the reactive 2.5PN orders, denoted 
by $\ddot { {\vek x}}_{1PN}$, $\ddot { {\vek x}}_{2PN}$ and 
$\ddot { {\vek x}}_{2.5PN}$ respectively, are non-spin by nature. The explicit expressions for these 
contributions, suitable for describing the 
binary black hole dynamics, were derived for the first time in the harmonic gauge 
\citep{TD}. These are given by 
\bea 
\ddot { {\vek x}}_{1PN} =& 
- {\frac{ G\,m }{c^2\, r^2}} \biggl\{  \vek n \left[ -2(2+\eta) \frac{G\,m }{r} \right.  &\\ \nn 
&\left. + (1+3\eta)v^2 - \frac32 \eta \dot r^2  \right] 
  -2(2-\eta) \dot r {\vek v} \biggr\} \,, \label{a1PN} & 
\\ \nn 
\ddot { {\vek x}}_{2PN} 
=& - \frac{G\,m }{c^4\, r^2} \biggl\{ {\vek n} \biggl[ \frac34 (12+29\eta)\left(\frac{G\, m}{ r}\right)^2&\\ \nn 
&+ \eta(3-4\eta)v^4 + \frac{15}{8} \eta(1-3\eta)\dot r^4 & \\ 
& - \frac{3}{2} \eta(3-4\eta)v^2 \dot r^2 - \frac12 \eta(13-4\eta) (\frac{G\,m}{r})\, v^2&\\ \nn 
&- (2+25\eta+2\eta^2) ( \frac{G\,m}{r})\, \dot r^2 \biggr] & \\ \nn 
& - \frac12 \dot r {\vek v} \left[ \eta(15+4\eta)v^2 - (4+41\eta+8\eta^2) (\frac{G\,m}{r})\right. &\\\nn 
&\left. -3\eta(3+2\eta) \dot r^2 \right] \biggr\} \,, \label{a2PN} & 
\eea\bea
\ddot { {\vek x}}_{2.5PN}  =& \frac{8}{15} \frac{ G^2 m^2 \eta }{ c^5 r^3 } 
\biggl \{ \left[ 9 v^2 + 17 \frac{G\,m}{r} \right] \dot{r} {\vek n}& \\  \nn 
&- \left[ 3 v^2 + 9 \frac{G\,m}{r} \right] { \vek v} 
\biggr \} 
\,,& 
\eea 
where the vectors $\hat {\vek n}$ and ${\vek v}$ are defined to be 
$ \hat {\vek n} \equiv {\vek x}/r $ and 
$ {\vek v} \equiv d {\vek x}/dt $, respectively, while 
$ \dot r \equiv dr/dt = \hat {\vek n} \cdot {\vek v}$, 
$ v \equiv | {\vek v} |$ and the symmetric mass ratio $\eta = m_1\, m_2/m^2$. 
The leading order spin-orbit contributions to $ \ddot {{\vek x} }$, 
appearing at 1.5PN order \citep{bar79}, reads 
\begin{eqnarray} 
\ddot { {\vek x}}_{SO} &= \frac{ G\, m}{ r^2} \, \left ( \frac{G\,m}{c^3\, r} \right )\, 
\left ( \frac{ 1 + \sqrt{1 -4\,\eta} }{4} \right )&\\ & \nn 
\, \chi 
\biggl \{ \biggl [ 12 \, \left [ {\vek s}_1 \cdot ( \hat {\vek n} \times {\vek v} ) \right ] \biggr ]\, 
\hat {\vek n} \\ & 
+ \biggl [ \left ( 9 + 3\, \sqrt {1 - 4\, \eta} \right ) \, \dot r \biggr ] 
\left ( \hat {\vek n} \times {\vek s}_1 \right ) 
\nn \\ & 
- \biggl [ 
7 +  \sqrt {1 - 4\, \eta} 
\biggr ] 
\left ( {\vek v} \times {\vek s}_1 \right ) 
\biggr \}\,,\nn 
\end{eqnarray} 
where the Kerr parameter $\chi$ and the unit vector 
${\vek s}_1$ define the spin of the primary black hole by the relation 
${\vek S}_1 = G\, m_1^2 \, \chi\, {\vek s}_1/c$ 
and $\chi$ is allowed to take values between $0$ and $1$ in general relativity. 
Further, the above expression for $\ddot { {\vek x}}_{SO}$ implies that the  
covariant spin supplementary condition is employed to define the center-of-mass world line of 
the spinning compact object in the underlying PN computation \citep{kid95}. 
Finally, the quadrupole-monopole interaction term $\ddot { {\vek x}}_Q $, entering at the 2PN order \citep{bar79}, reads
\begin{eqnarray} 
\ddot { {\vek x}}_Q & =- q \, \chi^2\, 
\frac{3\, G^3\, m_1^2 m}{2\, c^4\, r^4}
\, \biggl \{ 
\biggl [ 5(\vek n\cdot \vek s_1)^2 
-1 \biggr ] {\vek n}
\nn \\&
-2(\vek n\cdot \vek s_1) {\vek s_1} \biggr \}, 
\end{eqnarray} 
where the parameter $q$, whose value is $1$ in general relativity, is introduced to test the black hole `no-hair' 
theorem \citep{Thorne1980,WexKopeikin1999,Will2008}. 
The precessional motion for the 
primary black hole spin is dominated by 
the leading order general relativistic spin-orbit coupling 
and the relevant equation reads 
\begin{eqnarray}\nn
\frac{d {\vek s}_1}{dt} = {\vek \Omega} \times {\vek s}_1 \,,~~ ~ ~ ~ ~ ~ ~ ~ ~ ~ ~ ~ ~ ~  ~  ~  ~  ~  ~  ~  ~  ~   \\ 
{\vek \Omega} = \left ( \frac{G\,m\,\eta}{2c^2\, r^2} \right )  
\biggl ( \frac{ 7 + \sqrt {1 - 4\, \eta} }{ 1 + \sqrt {1 - 4\, \eta} } 
\biggr ) 
\left ( \hat {\vek n} \times \hat  {\vek v} \right ). 
\end{eqnarray} 
It should be noted that the precessional equation for the unit spin vector ${\vek s}_1$ enters the 
binary dynamics at 1PN order, while the spin contribution enters 
$ \ddot {{\vek x} }$ 
at the 1.5PN order. 

\begin{figure}[ht] 
\epsscale{1.0} 
\plotone{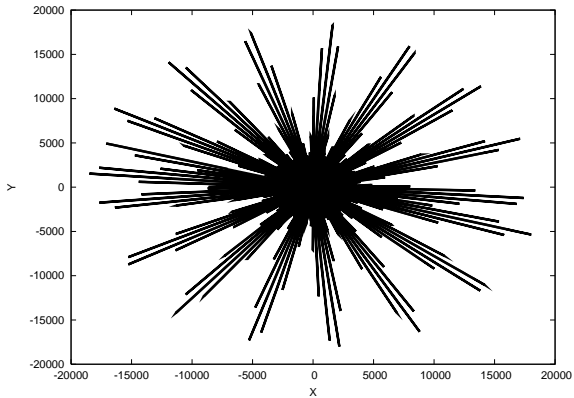} 
\caption{\label{fig3} 
  Precession of the orbital plane  due to the post-Newtonian spin-terms in a model 
in which the spin vector is kept constant (for clarity of the figure). 
The system is seen from the direction of the spin axis 
which in our model is in the orbital plane and this situation persists 
with the precision of approximately  half a degree when 
the plane itself precesses. 
    The time interval illustrated is 750 years during which the plane 
  precesses about $180^\circ$. At the same time the periastron 
  precesses much more rapidly and thus we see the orbit alternately 
  from the direction of aphelion, perihelion or sideways, which 
  causes the apparent different lengths of the small BH excursion 
  from the big BH (which is located at the origin of the figure).  } 
\end{figure}

\begin{figure}[ht] 
\epsscale{1.0} 
\plotone{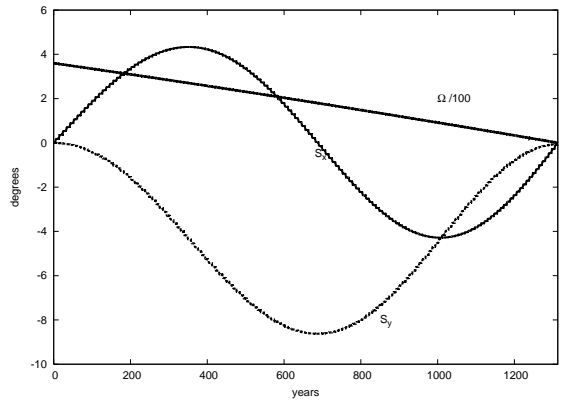} 
  \caption{
  The circulation of the ascending node ( $\Omega$) of the orbit 
   (straight line) in the coordinate system in which the initial 
   spin vector $\vek s_1$  points to the direction of the z-axis. 
   The quantities plotted are  $\Omega/100$ in degrees and 
   the symbols $s_x$ and $s_y$ correspond to the $x$ and $y$ 
   components of the spin-direction vector $\vek s_1$ in degrees.\label{fig4} 
   The plot for $\Omega$ thus illustrates (partly) the same behaviour as Fig.\ref{fig3}. 
} 
\end{figure}

  The main consequence of including the leading order spin-orbit interactions to the dynamics 
of a binary black hole is that it forces both the binary orbit and the orbital plane to precess. Moreover, 
the orbital angular momentum vector, characterising the orbital plane, precesses around 
the spin of the primary in such a way that the angle between the orbital plane and 
the spin vector ${\vek s}_1 $ remains almost constant (roughly within $\pm 0.\!^\circ5$ in our model). 
This is illustrated in the Figure \ref{fig3}. 
The spin-vector itself precesses drawing a cone with an opening angle of about 12 degrees 
(see Figure \ref{fig4}). 


The precessional period for both the orbital plane and the spin of the binary, provided by $|{\vek \Omega}|$, 
is about 2400 years, while employing the already inferred orbital parameters for OJ287 \citep{val07}. 
The relevant information required to evaluate $|{\vek \Omega}|$ are the primary mass $m_1 = 1.8\times10^{10}\, M_{\odot}$, 
the secondary mass $m_2 = 1.2\times10^{8}\, M_{\odot}$, the orbital period of 9 yr (without the redshift factor), 
and the eccentricity at apocenter $\sim 0.66$. Further, the orbital inclination 
relative to the plane of symmetry of the spinning black hole ( and presumably relative to the 
accretion disk of the primary as well) is taken to be 90 degrees. 
The manner in which the orbital plane precesses around ${\vek s}_1 $ is 
displayed in Figure \ref{fig3}. 
In the next section, we 
probe its consequences to the disk crossing timings. Contrary to the
illustration in Figure \ref{fig3}, the spin precession is taken into account in the timing experiments.

\section{Timing experiments}

In previous work (e.g. Valtonen 2007) the timing experiments were performed using the 1913, 1947, 1973, 1983, 1984 and 2005 outbursts as fixed points. They determine 5 time intervals whereby it is possible to determine uniquely 5 parameters. They do not include the secondary mass, and neither can we determine the Kerr parameter $\chi$ and the no-hair parameter $q$ without considering more fixed points.
\begin{figure}[ht]  
\epsscale{1.15}  
\plotone{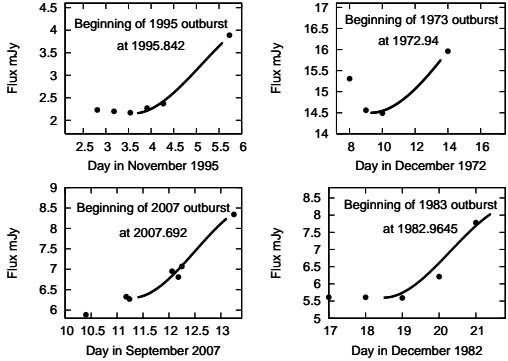}
\caption{The beginning of the best timed outbursts at 1973, 1983, 1995 and 2007. 
For 1973 and 1983 the points are daily averages, 
for 1995 averages over 8 hours, and single observations for 2007. 
The curve is a model fit for a spherical homogeneous bubble which 
becomes instantaneously transparent at the initial moment, given inside the figure.\label{fig5}}
\end{figure}  
Here we also make use of the 1995 and 2007 outbursts which have been observed with high time
resolution (see Figure \ref{fig5}) and they are therefore suitable for further refinement of the model. In addition, the newly discovered 1957 outburst is taken as another fixed point. The observed outburst times are listed in 
Table \ref{outburst}. Note that we have used the beginning of the
outburst as defined by the light curve rather than a linear extrapolation from the steepest rising slope.
This change makes the outburst times a little earlier than was previously assumed; e.g.
the 1983 outburst timing is now 1982.964 rather than 1983.00 (see Figure \ref{fig5}).
With this full set of outbursts no solutions were found unless the parameter $\chi$ is in the range 0.2 - 0.36. The
majority of solutions cluster around $\chi=0.29$, with a one standard deviation of about 0.04.

\begin{figure}[ht] 
\epsscale{1.15}  
\plotone{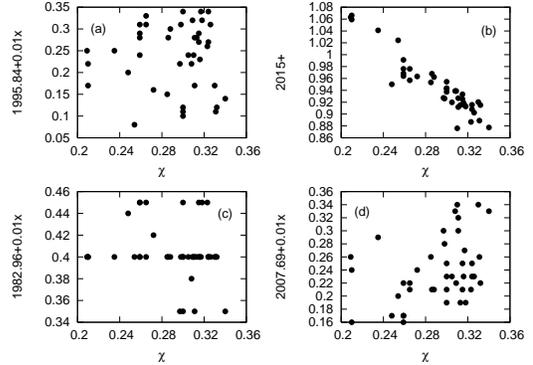}
\caption{Variation of the expected outburst timing as a function of black hole spin for four different outbursts. Results
from the various acceptable solutions are represented by points.\label{fig6}}
\end{figure}

Figure \ref{fig6} shows the distribution of points, each representing a solution, in four panels. Each panel displays the outburst time as a function of $\chi$. Panels (a), (c) and (d) are essentially scatter diagrams, 
demonstrating that the solutions cover the allowed range rather well. 
On the contrary, panel (b) shows a strong correlation between the time of the 2015 outburst and $\chi$. The range of possible outburst times extends from early November if $\chi=0.36$ to late January 2016 if $\chi=0.2$.

\begin{table}[t] 
\caption{Outburst times with  estimated uncertainties\label{outburst}. 
These are starting times of the outbursts.} 
\begin{tabular}{lr} 
1912.980& $\pm$  0.020\\ 
1947.283& $\pm$  0.002\\ 
1957.080& $\pm$  0.030\\ 
1972.945& $\pm$  0.012\\
1982.964&$\pm$   0.0006\\  
1984.130&$\pm$   0.005\\ 
1995.842&$\pm$   0.0015\\ 
2005.745& $\pm$  0.012\\ 
2007.692&$\pm$   0.0015\\ 
\end{tabular} 
\end{table} 
The timing experiments give a unique solution for the system parameters: \:\: 
precession in the orbital plane per period $\Delta \phi$, 
masses $m_1$ and $m_2$ of the two black holes, 
spin parameter $\chi$, initial phase $\phi_0$, initial apocenter eccentricity $e_0 $, 
`no-hair' parameter $q$ and 
time-delay parameter  $t_d$. The results are shown in Table \ref{solutiontable}. 

Here the precession rate is defined as the average change of the apocenter phase angle over the last 150 yrs.
The time delay parameter depends on the structure of the accretion disk, and obtains 
the value of unity in the model of 
\citet{leh96}. 
Its value is related to the thickness of the accretion disk, as explained in \cite{val06b}. 
\begin{table}[h] 
\caption{Solution parameters.\label{solutiontable}} 
\begin{tabular}{l|r} 
$\Delta \phi^\circ$  & $39.1 \pm 0.1$\\ 
$m_1$ & $(1.83\pm 0.01)\cdot 10^{10} M_\odot$ \\ 
$m_2$ & $(1.4\pm 0.1)\cdot 10^8 M_\odot$\\ 
$\chi$ &  $0.28\pm 0.08$\\ 
$\phi_0$ & $ 56\!^\circ.5\ \pm 1.\!^\circ2$\\ 
$e_0 $ & $0.6584\pm 0.001$\\ 
$q$ & $1.0\pm 0.3$\\ 
$t_d$& $0.75\pm 0.04$\\ 
\end{tabular} 
\end{table} 

Because we are using 8 time intervals to fit 8 parameters, the fit has no degrees of freedom. The tolerance limits of Table 4 are a consequence of having a tolerance in the fixed points of the outburst times (Table 3). We have carried out further experiments trying to evaluate how well determined these tolerance limits are. In case of $q$ values outside the range of Table 4 were found in a limited part of the initial value space. Thus the error limits for this quantity are likely to be one sigma error limits. For other variables we were not able to get values outside the range except by changing the astrophysical model, as reported below.

As part of the solution we obtain the list of all past and future outburst times with their uncertainties. For example, the well recognized outbursts in 1959, 1971 and 1994 are timed at $1959.213\pm0.002$, $1971.1265\pm0.002$
and $1994.6085\pm0.005$, 
respectively. In all these cases data are missing at the crucial time of expected rapid flux rise, 
and thus these predictions cannot be verified at present.

The value for the primary mass is within errors the same as was obtained by  \cite{leh96}, $m_1$ = $1.71\pm 0.15\cdot 10^{10} M_\odot$. Also the secondary mass is surprisingly close to the value of \cite{leh96},  $m_2$ = $1.36\cdot 10^{8} M_\odot$, when adjusted for the 'current' Hubble constant of 72 km/s/Mpc and for the 5.6 mJy outburst strength, as observed in 2007 \citep{val08b}. This mass value is based on the astrophysics of the impact and the strength of the maximum signal. The previous values of \cite{leh96} were 80 km/s/Mpc and 5.1 mJy which correspond to the secondary black hole mass $1.0\cdot 10^{8} M_\odot$. The updated astrophysically determined secondary mass is within the error limits of the dynamically determined mass. Considering that the two ways of measuring the mass are completely independent, the agreement is a good indicator of the correctness of the basic model. 

We may ask what is the likelihood of finding a quasar like OJ287 in an all sky search for periodic quasars. One may approach this issue from a theoretical point of view in cosmological evolution models, or as an observational problem. In the former case it has been calculated that many binary black holes are expected in quasars, e.g. Volonteri et al. (2009) estimate that there should be about $10^{4}$ sub-parsec binaries up to redshift $z=0.7$ among quasars with absolute $i$-magnitude less than -22, and mass ratio not exceeding 100 over the whole sky. For a system like OJ287 there is something like 25$\%$ chance to be found at a redshift below 1 in cosmological simulations (private communication by an anonymous referee).

When considering the observational situation, we note that at its faintest OJ287 has been $m_B=18$; since the jet of OJ287 is strongly beamed toward us, we use this base magnitude in comparing with other quasars. There are about $2\cdot 10^{4}$ quasars brighter than this limit in the sky (Arp 1981). Many of these quasars may host binary black holes (Comerford et al. 2009). Thus potentially there are about $10^{4}$ bright (at OJ287 base level) binary quasars in the sky. The rate of binaries in quasars is somewhat uncertain at present.

Let us again consider the special case of OJ287. Vestergaard $\&$ Kelly (private communication) estimate that there is about a unit probability of finding a broad-line quasar brighter than $i$-magnitude 18 where the black hole mass exceeds $10^{10}M_\odot$ in a search covering the whole sky. The probability of finding one hundred such quasars is about 25$\%$.  This is an extrapolation from the study of SDSS (DR3) quasars (Vestergaard et al. 2008), complemented by the BQS sample. The effect of the bright limit ($i=15$) in the SDSS catalog which reduces the number of observed nearby massive quasar black holes, has been corrected for. Altogether, the estimate based on observations is not inconsistent with the estimate from the cosmological evolution models, considering the large uncertainties in both.

According to our model, OJ287 is in the stage of final inspiral before the merger of the two black holes. Its remaining lifetime of about $10^{4}$ yr is only a fraction $10^{-3}$ of the expected time of black hole binary evolution of about $10^{7}$ yr (Volonteri et al. 2009). On the other hand, OJ287 is strongly beamed toward us. Even though the likelihood that we happen to be looking straight at the beam of a quasar is only $\gamma^{-2}/4$ where $\gamma$ is the Lorenz factor of the beam, the intensity of the source is magnified by $\gamma^3$, and the probability of detection in a flux limited sample goes as the 1.5-power of the intensity. The joint probability that OJ287 enters the samples of both optically bright and radio loud quasars (the selection criteria for the Tuorla monitoring program) is raised by several orders of magnitude above the corresponding probability for unbeamed quasars. At its brightest, OJ287 is about 5 magnitudes brighter than at its minimum light. This increase in detection probability counters the decrease in detection probability due to the limited lifetime of its currect stage of evolution. Based on these considerations, it is not surprising that a short period binary quasar with the primary heavier than $10^{10}M_\odot$ has been discovered.
 
There is nothing special about other parameters of OJ287 either. The large mass ratio is necessary for the stability of the accretion disk. The moderately high eccentricity is what is expected at the OJ287 stage of inspiral of binary black holes of large mass ratio (Baumgardt et al. 2006, Matsubayashi et al. 2007, Sesana et al. 2008, Iwasawa et al. 2009). The value for $t_{d}$ implies $\alpha\sim0.1$ in the standard disk model, an acceptable value. In the next section we will discuss the main targets of this paper, the values of $\chi$ and $q$.

We have performed numerical experiments to study the stability of the above listed solution.
In the first instance, we added the conservative, non-spinning 3PN corrections to
$\ddot { {\vek x}} $  describing the relativistic binary black hole orbital dynamics \citep{mw04}.
The timing experiments employing the improved  binary black hole  dynamics lead to
orbital solutions that are essentially close to what is listed in Table 3.
The only noticable difference is the roughly $1.5 \%$ increase in the estimated mass of the
primary black hole. This result suggests that
the employed PN-dynamics, relevant for our  binary black hole model,  is in the convergent
regime. The increase in the estimated mass of the primary can be qualitatively explained
by evaluating the period of periastron precession $T_{\rm prec}$ at 2PN and 3PN orders
with the help of Eqs.~(20) in \cite{mgs04}.
Using a non-spinning binary black hole model for OJ287, we obtain $T_{\rm prec} \sim 87$ years at the 2PN level
and an increase in the primary mass by $1.5 \%$ reduces it to $\sim 86$ years.
The similar estimate involving the
non-spinning binary black hole model
at the 3PN order gives $T_{\rm prec} \sim 83$ years indicating that
our timing experiments are qualitatively consistent with our rough theoretical estimates for $T_{\rm prec}$.
The quantitative differences may be attributed to the fact that in the timing experiments, we employed
spinning binary black hole model that decays under gravitational radiation reaction.
We would like to point out that it is the data set
spanning roughly 100 years that makes
$T_{\rm prec}$ a good diagnostic tool in analyzing our timing experiments.
Let us also note that at present the 3.5PN reactive contributions to $\ddot { {\vek x}} $
do not influence our estimates in any siginificant way. This is because its contribution to
the rate of change of the binary black hole orbital period is $\sim 0.00014$ and this is within the error limits of
our analysis.

The second test involves varying the astrophysical parameters of the model. The free parameter $t_{d}$ already tests the different acceretion disk structures. Any errors in the model for the evolution of the radiating bubble affect primarily this parameter, not others. There are two more astrophysical parameters which have fixed values in our model; here we will briefly discuss their influence on the solution. The disk level above or below the mean level is affected by the tides raised by the secondary. The effect of tides on the impact times was calculated by Valtonen (2007), and it is described by an analytical function of the impact distance. We have tested possible errors in this formula by varying the disk tidal level by a constant factor. It turns out that the level can vary anywhere between $40\%$ and $200\%$ of the standard value without affecting our result. No solutions were found if the disk level correction is outside this range. Another fixed parameter in the model is related to the asymmetry of the impact with respect to the central plane of the disk. First the secondary impacts the upper surface of the disk, initiates the evolution of the radiating bubble, and then one disk crossing time later it impacts the lower surface of the disk and initiates the evolution of the second radiating bubble (Ivanov et al. 1998). Since we observe the disk from one side only, we see one or the other of the two bubbles. Thus our model must include the delay by one crossing time when we observe the bubble from the 'lower' level. We derive the value $1.05\pm0.15$ crossing times for the delay from the illustrations of Ivanov et al. (1998). Since in the double impacts the radial speed changes its sign relative to us between the approaching part of the orbit and the receding part of the orbit, every second outburst must include the one crossing time delay. What happens if the delay is not exactly one crossing time? We have varied the delay by a constant factor, and found that if the delay is within the range of $0.85 - 1.20$ crossing times, it does not affect our result. It is unlikely that the delay would in fact fall outside this range. However, if the delay is for some unknown reason much shorter than one crossing time, solutions are found down to the level of $0.55$ of the crossing time, but then the spin value is greater than in our standard model. At the extreme, the spin can be as high as $\chi=0.43$. As the correlation of Figure 6b shows, this possibility can be eliminated (or verified) in 2015. The same correlation still holds even if the delay is much less than unity.

\section{Conclusions} 

  The binary black hole in OJ287 is modelled to contain a spinning primary black hole with an accretion disk
  and a non-spinning secondary black hole. Using PN-accurate dynamics, relevant for such a system, 
we infer that the primary black hole 
should spin approximately at one quarter of the maximum spin rate allowed in general relativity. 
In addition, the `no-hair theorem' of black holes
\citep{MisnerThorneWheeler1973}
is supported by our model, although the testing
is possible only to limited precision. 
The solutions concentrate around $q=1.0$ with one standard deviation of $0.3$ units.
These results are achieved with the help of new data on historical outbursts as well 
as using the most recent outburst light curves together with 
the timing model for OJ287 outbursts \citep{val07}. 
To obtain the above estimate for the Kerr parameter $\chi$ and the no-hair parameter $q$, 
we have assumed that the spin of the primary black hole is initially (150 yrs ago) 
aligned with respect to the spin of accretion disk. 
Further, a polar orbit is assumed as in previous work.
The main justification for the polar orbit 
is the sudden fade of OJ287 over a wide range of wavelengths in 1989, 
presumably because of an eclipse of the jet by the secondary \citep{Takalo_et_al.1990}. 
The polar orbit was also used to justify the rather modest rise of flux in 1995, 
possibly because the secondary moved parallel to the disk axis after the disk impact 
and blocked off some of the optical emission  \citep{val06b}. 

What happens if these conditions are not satisfied? Sundelius et al. (1997) carried out simulations with a number of different inclinations between the disk and the orbit, and found that the inclination makes no difference. This is not surprising since the impacts occur along the line of nodes, and the timing of the impacts is not a function of the impact angle. In this work we have tried varying the spin angle relative to the disk, and found that within reasonable limits (we considered inclinations up to $10^\circ$) the results do not change. Again this is what is expected since the line of nodes circulates slowly in comparison with the orbital angular motion.

The reason why we would expect at least approximate alignment between the black hole spin and the disk spin is the Bardeen - Peterson effect which tends to align the two spins. However, the alignment time scale is much slower (about $10^{7}$ yr, Lodato \& Pringle 2006) than the spin precession time (Fig. 4), and thus the disk does not follow the black hole spin exactly but only on average.

In this scenario, we have a unique solution and also a unique prediction for the next OJ287 outburst, 
expected in 2015.  
We should then be able to judge the correctness of the present solution. Note
that an outburst is not expected in 2015 in any simple extrapolation from past observations, as it is well before the the
average 12 yr cycle is due, and thus it is a sensitive test of the general model as well as a test 
for the spin of the primary black hole.

In the meantime, there is an additional observation which supports the spin value that has been derived here. OJ287 has a
basic 46$\pm$3 day periodicity \citep{Wu06}, which may be related to the innermost stable orbit in the accretion disk. 
However, since we presumably
observe the accretion disk almost face on, and there is an $m=2$ mode wave disturbance in the disk, this is likely to refer
to one half of the period. Considering also the redshift of the system, and the
primary mass value given in Table 4 this corresponds to the spin of $\chi= 0.35 \pm0.06$ \citep{McClintocketal}. The uncertainty of 0.06 units is related to the
width of the trough in the structure function of the flux variations. 

It will be interesting to explore the consequence of allowing the secondary black hole to spin. 
This will require 
including the leading order spin-spin interactions, appearing at the 2PN order, to the 
inspiral dynamics \citep{kid95}. 
We are also planning to probe the effect the higher order 
contributions to the spin-orbit coupling, entering 
the binary black hole binary dynamics at the 2.5PN order \citep{BBF}. 

There are at least 5 additional outbursts in the historical record which have not yet
been been accurately observed. If new data of these outbursts are found it will open up the possibility
of improving the model and studying higher order effects. Searches of plate archives are planned for
this purpose.

\acknowledgments

We are grateful to Gerhard Sch\"afer and Cliff Will for helpful discussions on the dynamics of 
spinning black hole binaries. We also thank Marianne Vestergaard, Brandon Kelly and Masaki Iwasawa for information prior to publication, and the anonymous referee for providing results of a calculation, as well as for other suggestions for improving the paper.
AG is partly supported by 
DFG's SFB/TR 7 ``Gravitational Wave Astronomy'' 
and DLR (Deutsches Zentrum f\"ur Luft- und Raumfahrt). RH acknowledges the Grant Agency of the Czech Republic, grants 102/09/0997 and 205/08/1207, and MSMT ME0927.
MB and RH acknowledges partial support by the Smithsonian Institution/CfA.

\end{document}